\documentstyle[psfig]{qcdparis}
\newcommand{\ket}[1]{\mbox{$| #1 >$}}
\newcommand{\bra}[1]{\mbox{$< #1 |$}}

\begin{document}
\pagestyle{plain}
\title{ ELFE : an Electron Laboratory for Europe. }
\author{
Christian Cavata$^a$ and Bernard Pire$^b$
}

\affil{
a) DAPNIA, Centre d'Etudes Nucl\'eaires de Saclay,
 91191 Gif sur Yvette, France.\\
b) Centre de Physique Th\'eorique, Ecole Polytechnique,
 F-91128 Palaiseau, France.
}

\resume{
Cet article pr\'esente un bref survol du  domaine de  physique
pour lequel la communaut\'e  europ\'eenne de physique nucl\'eaire  vient
de recommander de construire un acc\'el\'erateur d'\'electrons de 15 \`a
30~GeV \`a faisceau  continu.  La question  centrale   \'etudi\'ee gr\^ace
\`a cette machine sera  la structure en quarks et  gluons des
hadrons.
}

\abstract{
{This paper presents a brief overview of the physics with  the
15-30~GeV continuous beam electron facility proposed by the European
community of nuclear physicists  to study the quark and gluon structure of
hadrons.}
}

\twocolumn[\maketitle]
\fnm{7}{Talk given in the
"Future Prospects" session
at the Workshop on Deep Inelastic scattering and QCD,
Paris, April 1995 and at the VIth International Conference on Elastic
and Diffractive Scattering, Blois, June 1995}

\section{Introduction}

Recently,  the Nuclear Physics European  Collaboration Committee (NuPECC)
of the European  Science
Foundation  has  recommended~\cite{NUPE95}  the  construction  of  a  15-30
GeV  high
intensity continuous beam  electron accelerator.   The goal of  this new
facility  is  to  explore  the  quark  structure  of matter by exclusive
and semi-inclusive electron scattering from nuclear targets.

In the last  two decades, we  have seen the  emergence of a  theory that
identifies the  basic constituents  of matter  and describes  the strong
interaction~\cite{QCD20}.    The  elementary  building blocks of atomic
nuclei  are  colored  quarks  and  gluons.   The theory describing their
interactions  is  Quantum  Chromodynamics  (QCD)  which  has two special
features, asymptotic freedom and color confinement.  Asymptotic  freedom
means  that  color  interactions  are  weak  at  short distances. At  large
distances, color confinement results in the existence of hadrons and in the
 impossibility to observe quarks and gluons as single particles.
\bigskip

Although one knows the  microscopic theory for the  strong interactions,
{\em one does not understand how quarks build up hadrons}.

\section{\bf The ELFE project}

The ELFE  research  program\cite{ELFE},\cite{ARPI95} lies  at  the  border
  of  nuclear and particle
physics. Most  of the predictions  of QCD are  only valid at  very high
energies  where  perturbation  theory  can  be  applied.
Understanding however how  hadrons  are  built,
is  the domain of
confinement where  the coupling  is strong.   Up  to now  there are only
crude theoretical  models of  hadronic structure  inspired by  QCD.  One
hopes that in the next  ten years major developments of  nonperturbative
theoretical methods such as lattice gauge theory will bring a wealth  of
results on the transition  from quark to hadron.   However, many theorists
think that it is fundamental  to
guide   theory   with   accurate,   quantitative  and  interpretable
measurements obtained by electron scattering experiments.

The research program of ELFE addresses the questions raised by the quark
structure of matter: the role  of quark exchange,  color transparency,
flavor  and  spin  dependence  of  structure  functions  and differences
between  quark   distributions  in   the  nucleon   and  nuclei,   color
neutralization in the hadronization of a quark\ldots All these questions
are some of the many exciting facets of the fundamental question:

\begin{center}
{\bf ``How do color forces build up hadrons from quarks and gluons? ''}
\end{center}
\noindent
ELFE will focus on the following research topics:
\begin{itemize}
\item {\bf Hadron structure} as revealed by  hard exclusive reactions :
 baryon form factors,   real and virtual Compton scattering,
electro and photo-production of mesons ($\pi$,$K$,$\rho$,$\phi$ ...),
 meson ($\pi$,$K$, ...) form factors.
\item{\bf Evolution from  mini-hadron to hadron} in Color Transparency
experiments.
\item {\bf Vector mesons and heavy quarks} : diffractive production and
exclusive
 scattering at high transfers.
\item {\bf Space time picture of quark hadronisation} through the study of
absorption by
 nuclear medium.
\item {\bf Separation of  Valence and Sea } content of the proton
  by tagged structure functions measurements.
\item {\bf Study of  spin structure } of the nucleon through semi-inclusive
experiments.
\item {\bf Light nuclei short distance structure } through form factor
measurements and
deep inelastic scattering at $x>1$.
\end{itemize}
We will here concentrate on exclusive reactions. The concept and
 experimental situation of color transparency have been presented elsewhere
\cite{PIRE95}. Hadronization and tagged structure functions are already
 classical fields of QCD, but they should be an important part of the ELFE
program
 because of the specific parameters foreseen for the machine.

\section{\bf Hard Exclusive reactions:  A new tool}

\begin{figure}
\begin{center}
\begin{tabular}{ccc}
\psfig{file=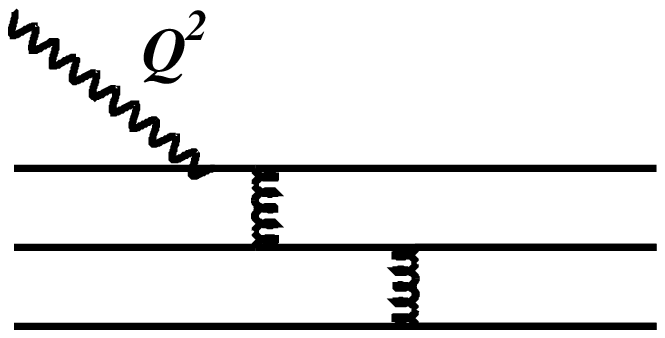,width=3cm} &
\hspace{-1cm} \psfig{file=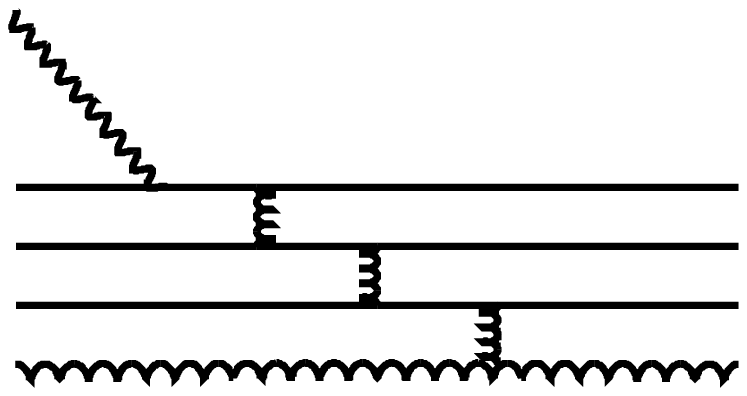,width=3cm} &
\hspace{-1cm} \psfig{file=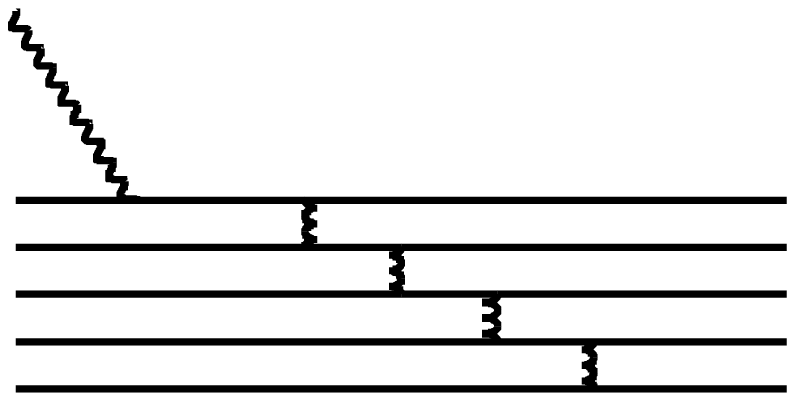,width=3cm}
\end{tabular}
\caption{Hard scattering amplitudes for the proton form factor.
1/Q$^2$ expansion\label{TH} }
\end{center}
\end{figure}

Exclusive reactions\cite{SUPERPIRE} are processes in which the final state
is completely
resolved.  They  are important since  at high momentum  transfers
they select the simplest non perturbative objects. This is what we will now
explain.

\subsection{ Factorization and Valence selection}
Perturbative QCD studies have shown that factorization properties allow to
 separate well-defined non perturbative
objects which are crucial for the understanding of confinement dynamics from
perturbatively calculable hard processes.

\begin{figure}
\begin{center}
\mbox{ \psfig{file=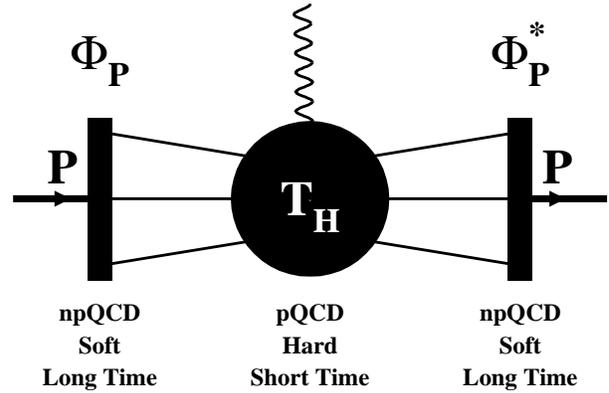,width=8cm} }
\caption{The factorization of a hard scattering amplitude \label{hard_bw} }
\end{center}
\end{figure}

One starts with the Fock expansion\cite{EXCLTHEORY} with a fixed number of
quarks and gluons for a proton state of momentum $P$ :
$$
\ket{P} =
\Psi^P_{qqq} \ket{qqq} +
\Psi^P_{qqq,g} \ket{qqq,g} +
\Psi^P_{qqq,q\bar{q}} \ket{qqq,q\bar{q}} +
 ...
$$
where
$
\psi^P_l(x_i,\vec{k}^{\perp}_{i})$
describes how the $l$ quarks and gluons share the proton momentum.
The    wave  functions  $\Psi^P_l$  are  functions  of light-cone
momentum fraction $x_i$, transverse momentum $k^\perp_i$ and helicities.
They contain  the  information  on  quark  confinement  dynamics.
Here the quarks are ``current'' quarks and not ``constituent'' ones.

  A
constituent quark  may be  seen as  a complex  structure consisting of a
current  quark   ``dressed''  of   quark-antiquark  pairs   and  gluons.
Constituent quarks  are important  to get  an intuitive  picture of  the
quark structure of  the nucleon, but  they cannot be  used to understand
quark dynamics in the framework of a relativistic quantum field theory.

Of particular interest is
$
\Psi^P_{qqq}
$,
 the valence proton wave function. This is the simplest non perturbative object
 in the proton.

Let us take the example of the proton form factor. As depicted on
figure~\ref{TH},
 the ``three quarks'' hard scattering amplitude gives a contribution to the
form
factor proportional to $\left[1/{Q^2}\right]^2$ due to the two gluon
 propagators, whereas the ``3~quarks~1~gluon'' amplitude, requiring three gluon
 propagators, contributes for
$\left[1/{Q^2}\right]^3$. The argument may be repeated for more
 participating constituents and for any reactions.
Thus, the valence component $\phi^P_{qqq}$ turns out to be the dominant one in
hard exclusive reactions.

Factorization is then the statement
 that a hard matrix element can be written as (see Figure \ref{hard_bw})
\begin{equation}
\bra{P} T \ket{P}\simeq \phi^{*P}_{qqq}  \otimes T_H^{qqq,qqq}
\otimes\phi^P_{qqq},\label{LEADING}
\end{equation}
up to $1/Q^2$ corrections. Integrals  over momentum fractions  $x_i$ and $y_j$
are implicit.
Here
$$\phi_{qqq}(x_i) ={\int} [dk^\perp] \psi_{qqq}(x_i,\vec{k}^{\perp}_{i})
$$
is the proton   Valence Distribution Amplitude,
and $T_H^{qqq,qqq}$ is a hard scattering amplitude, calculable in perturbative
QCD.

\begin{figure}
\begin{center}
\mbox{ \psfig{file=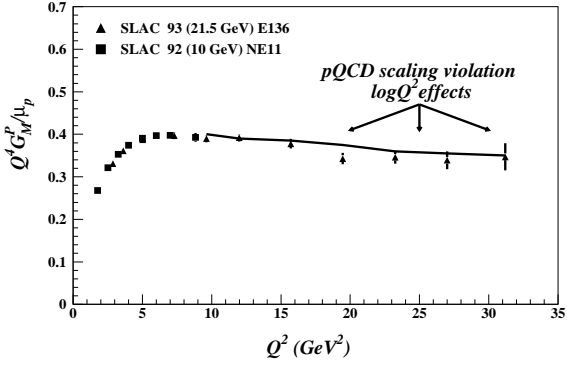,width=8cm} }
\caption{$Q^2$ evolution of the proton form factor. Above 10 GeV$^2$,
Scaling is established.  \label{gpm_slac} }
\end{center}
\end{figure}

The applicability of this factorization
in a definite energy domain is indicated by some definite statements, such as
the
logarithmically corrected dimensional counting rules, the helicity conservation
 law and the appearance of color transparency.
 The few data available~\cite{gpm_slac} (see figure \ref{gpm_slac}) indicate
that the ELFE parameters indeed
 correspond to this well defined physics domain. Nevertheless, checking
 factorization will be a necessary prelude of the experimental program at ELFE.

 Configurations of small transverse extension
are also selected by hard exclusive reactions. Indeed, in the Breit frame where
the
virtual photon
is collinear to the incoming proton which flips its momentum, the first
hit quark
changes its direction and gets a momentum $O(Q)$; it must transmit this
information
 to its comovers within its light cone; this can only be achieved if the
transverse
 separation is smaller than $O(1/Q)$. This is  the basis of the Color
Transparency
phenomenon.

\subsection{QCD evolution of wave functions}
The analysis of QCD  radiative corrections to any exclusive amplitude  has
shown~\cite{EXCLTHEORY}  that the factorized distribution  amplitudes  obey  a
renormalization group equation, leading  to a well understood  evolution
in terms  of perturbative  QCD.   At asymptotic  $Q^2$, the distribution
amplitudes simplify, e.g. for the proton valence distribution amplitude:
\begin{equation}
\phi^P_{qqq}(x_1,x_2,x_3,Q^2)\rightarrow 120 x_1x_2x_3\delta (1-x_1-x_2-x_3)
\end{equation}
\noindent
The $Q^2$ evolution  is however sufficiently  slow for the  distribution
amplitude  to  retain  much   information  at  measurable  energies   on
confinement physics.  The experimental strategy of ELFE physics is  thus
to sort out  the hadron distribution  amplitudes from various  exclusive
reactions to learn about the dynamics of confinement.

At finite $Q^2$  the proton valence wave-function can be written as a series
 derived from the leading logarithmic analysis,  in terms of
 Appel polynomials $P_i(x_i)$, as~\cite{EXCLTHEORY}
\begin{equation}
\label{APPEL}
\begin{array}{c}
\phi^P_{qqq}(x_i,Q^2)=120 x_1 x_2 x_3
\left\{\right.
1 +
{{21}\over{2}}
\left({\alpha_S(Q^2)\over\alpha_S(Q_0^2)}\right)^{\lambda_1} A_1 P_1(x_i)\\
+{{7}\over{2}}
\left({\alpha_S(Q^2)\over\alpha_S(Q_0^2)}\right)^{\lambda_2} A_2 P_2(x_i)
+...\left.\right\}.
\end{array}
\end{equation}
The unknown coefficients $A_i$ are governed by confinement dynamics.
\subsection{A strategy for data analysis}
A first way to analyse data is to compare experimental points to a calculation
with a given distribution amplitude that any prejudiced theorist convinced you
to choose.

A more model-independent way is to try to sort out the wave function
directly from the data. Let us outline a possible strategy on the example
 of real Compton scattering. Using the expansion of Eq.~\ref{APPEL},
we write the Compton differential cross-section as a sum of terms
\[
A_i T_H^{ij}(\theta) A_j
\]
\noindent
where $T_H^{ij}$ are integrals of the hard amplitude at some given
scattering angle $\theta$ multiplied by the two Apple polynomials
$A_i(x)$ and $A_j(y)$ integrated over light-cone variables $x$ and $y$. The
analytical expression is complicated but can easily be electronically
 managed.

Sorting out the valence wave function of the proton from the data
amounts then to determine through a maximum of likelyhood method the
parameters $A_i$ restricting to something like ten terms in the expansion
of Eq. \ref{APPEL}. A direct test of the validity of the approach is then to
 explore both real and virtual Compton scattering data at all angles
 which should be understood with the same series of  $A_i$'s.

\subsection{ Other processes }

Photo- and electro-production of mesons at large angle will enable to
probe $\pi$ and $\rho$ distribution amplitudes in  the same way. The
production of $ K \Lambda $ final states will enable to explore strange
 quark production, for which the diagrams in the hard process are more
restricted. Not
much theoretical analysis of these possibilities has however been worked
out except under the simplifying assumptions of the diquark
model~\cite{DIQUARK},\cite{FARA91}.
\subsection{Deep Inelastic Scattering and Nucleon Wave function}
\begin{figure}
\begin{center}
\begin{tabular}{cc}
\psfig{file=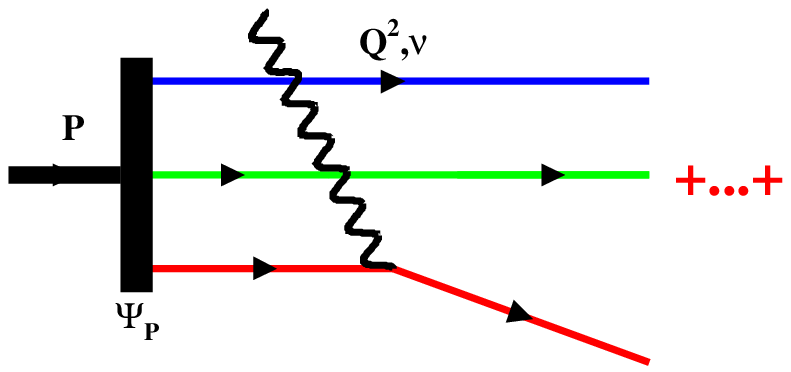,width=4cm}&
\hskip -4mm\psfig{file=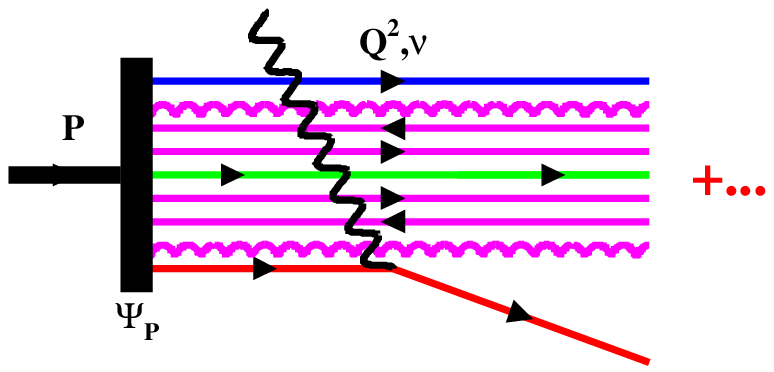,width=4cm}\\
\end{tabular}
\caption{Proton structure as seen by DIS experiments\label{DIS}}
\end{center}
\end{figure}

Nearly all  existing data  on quark  distributions in  hadrons have been
obtained by {\em \ inclusive  scattering} of high energy particles.   In
such reactions, one strikes quarks with considerable momentum and energy
and  reconstructs  quark  distributions  from  scattering data.  This is
possible  due to  the factorization.  This property has given a  firm basis
for the  partonic description, thereby allowing to  go beyond the
original  model  proposed  by  Feynman  and  Bjorken. Inclusive
scattering  amounts  to  an  average  over  all  the   possible  quark
configurations in the nucleon (Figure \ref{DIS}).  Indeed, the extraction of
structure functions
in DIS experiments gives access to parton densities $q_{a/P}$ of type $a$ in a
proton.
For instance the $F_2$ structure functions is
$$F_2(x,Q^2)=\sum_a e_a^2x q_{a/P}(x,Q^2).$$
In terms of wave functions, these "parton densities" are
$$ q_{a/P}(x,Q^2)=\sum_l\int[dx][dk^\perp]
|\Psi_l(x_i,k^\perp_i)|^2\delta (x_a-x)
$$
For instance for the $u$ quark,
$$
\begin{array}{llllll}
u_{a/P}=&\int[dx][dk^\perp]&|\Psi_{uud}|^2         &\delta (x_u-x)&+&\\
               &\int[dx][dk^\perp]&|\Psi_{uudg}|^2        &\delta (x_u-x)&+& \\
               &\int[dx][dk^\perp]&|\Psi_{uudgq\bar{q}}|^2&\delta (x_u-x)&+&...
\end{array}
$$
By contrast with exclusive reactions, there is no valence wave function
selection.
In addition to fundamental tests of QCD,
the measurements of  structure functions have  lead to the  discovery of
the importance of gluons in  the momentum and spin distributions  in the
proton. We  now  need  to  go  beyond  and  to understand  how simple quark
configurations  are  controlled  by  confining  mechanisms.  One needs a
different type of data  sensitive to the time  evolution of a system  of
correlated quarks.   This  is the  domain of  exclusive reactions  where
scattered  particles  emitted  in  a  specific  channel  are observed in
coincidence.

Due to the  smallness
of exclusive amplitudes at  large transfers,
existing  high  energy   electron
accelerators, designed to study electroweak physics, cannot  give access to
these
distribution amplitudes.  The only  way   to  study  exclusive  reactions  at
large
transfer is  to use  a dedicated  high intensity continuous
beam  accelerator.

\section{Accelerator and detectors}

The choice of the energy range of 15 to 30 GeV for the ELFE  accelerator
is fixed by a compromise between

\begin{itemize}

\item Hard electron-quark scattering:   one must have  sufficiently high
energy  and  momentum  transfer  to  describe  the  reaction in terms of
electron-quark scattering.  The high  energy corresponds to a very  fast
process where the struck quark  is quasi-free.  High momentum  transfers
are necessary to probe short distances.

\item  The smallness of the exclusive cross sections when the energy increases,
as exemplified
by the quark counting rules~\cite{EXCLTHEORY}.
 \end{itemize}

\begin{table}[hb]
\begin{center}
\begin{tabular}{|l|r|}
\hline\hline
Beam Energy & $15 \div 30 GeV$ \\
Energy Resolution FWHM & $3 \times 10^{-4}$ @ 15 GeV \\
& $10^{-3}$ @ 30 GeV \\
Duty Factor & $\simeq 100$~\% \\
Beam Current & $10 \div 50 \mu$A \\
Polarized Beams & $P > 80$~\% \\ \hline\hline
\end{tabular}
\end{center}
\caption{ELFE Accelerator Parameters}
\label{t31}
\end{table}

Exclusive and semi-inclusive  experiments are at  the heart of  the ELFE
project.  To avoid a prohibitively large number of accidental coincident
events a high duty cycle  is imperative.  The ELFE  experimental program
also  requires  a  high   luminosity  because  of  the   relatively  low
probability of exclusive processes.  Finally a good energy resolution is
necessary to identify specific reaction channels.  A typical  experiment
at 15  GeV (quasielastic  scattering for  instance) needs  a beam energy
resolution of about  5 MeV. At  30 GeV the  proposed experiments require
only  to  separate  pion  emission.    These characteristics of the ELFE
accelerator are summarized in table \ref{t31}.
\begin{figure}
\begin{center}
\mbox{\psfig{file=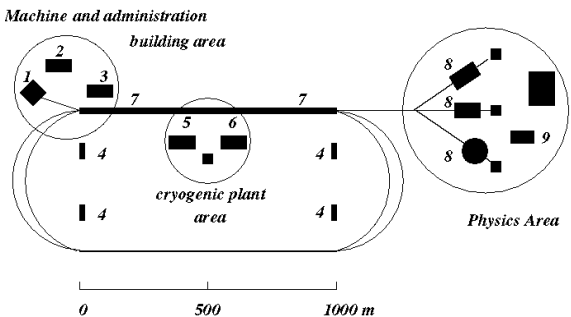,width=10cm}}
\vskip -20mm
\caption{ ELFE original design\label{SCHEME1} }
\end{center}
\end{figure}

Due  to  the  very  low  duty  cycle  available at SLAC and HERA (HERMES
program)  one  can  only  perform  with  these  accelerators   inclusive
experiments and a very limited set of exclusive experiments.

\begin{center}

{\em ELFE  will be  the first  high energy  electron beam  beyond 10 GeV
\\with both high intensity and high duty factor.}

\end{center}

 The design proposed in 1992
 for the machine\cite{ELFE} consists of a 1 km superconducting 5 GeV linac,
with three recirculations (figure \ref{SCHEME1}). Taking into  account the
improvement of cavity
 performances, and running at 70 \% duty cycle would result in a price of order
 200 MECU. Different designs could also be considered as for instance, a
solution which would combine
 a test  linac of 30 GeV with 1 \% duty cycle for the future e$^+$e$^-$
 collider (TESLA) and the existing  HERA ring for
stretching
the pulse.
The  various  components  of  the  ELFE experimental physics program put
different requirements on  the detection systems  that can be  satisfied
only  by  a  set  of  complementary  experimental  equipment.   The most
relevant detector features are  the acceptable luminosity, the  particle
multiplicity, the angular acceptance and the momentum resolution.   High
momentum   resolution   ($5   \times   10^{-4}$)   and  high  luminosity
($10^{38}$~nucleons/cm$^2$/s)  can  be  achieved  by  magnetic  focusing
spectrometers.   For semi-exclusive  or exclusive  experiments with more
than two  particles in  the final  state, the  largest possible  angular
acceptance  ($\sim  4  \pi$)  is  highly  desirable.    The  quality and
reliability of large acceptance detectors have improved substantially in
the last two decades.  The design of the ELFE large acceptance detectors
uses state of  the art developments  to achieve good  resolution and the
highest possible luminosity.

\section{CONCLUSIONS}

The ELFE  research program  lies at  the border  of nuclear and particle
physics.  Most  of the predictions  of QCD are  only valid at  very high
energies  where  perturbation  theory  can  be  applied.    In  order to
understand  how  hadrons  are  built,  however,  one has to go in the domain of
confinement where the  coupling is strong.   It is  fundamental to guide
theory  by  the  accurate,  quantitative  and interpretable measurements
obtained by electron scattering experiments, in particular in  exclusive
reactions.

 This research domain is  essentially a virgin territory.   There are
only scarce experimental data with poor statistics. This  lack
  of  data  explains  to  a  large  extent  the  slow  pace of
theoretical progress.   The situation can  considerably improve due  to
technical breakthroughs in electron accelerating techniques.  We believe
that future significant progress  in the understanding  of the
evolution from quarks to hadrons will be triggered by new information coming
from dedicated machines such as the ELFE project.

The goal of the ELFE research program, starting from the QCD  framework,
is to explore the coherent and quark confining QCD mechanisms underlying
the strong force.  It is not to test QCD in its perturbative regime, but
rather to use  the existing knowledge  of perturbative QCD  to determine
the reaction mechanism and access the hadron structure.

\vspace*{0.5cm} \noindent

{\em ELFE will use the tools that have been forged by twenty years of
research in QCD, to elucidate the central problem of color interaction:
color confinement and the quark and gluon structure of matter.}

\begin{center}
{\large\bf Aknowledgements}
\end{center}
It is a pleasure to thank the convenors of the
``Future Prospects'' working group and the organizing committee.
We would like to acknowledge the many lively discussions
on Elfe with many of our colleagues and especially Jacques Arvieux, Bernard
Bonin, Bernard Frois, Thierry Gousset, Pierre Guichon, Jean-Marc Laget,
Thierry Pussieux and John P.Ralston.
Centre de Physique Th\'eorique is Unit\'e propre du CNRS.

\end{document}